\documentclass[aps,prl,preprint,showpacs,nobibnotes,floatfix,legal]{revtex4}

\usepackage[latin1]{inputenc}                    % Codepage latin1
\usepackage{graphicx}                            % Graphics package
\usepackage{latexsym}                            % Load additional symbols
\usepackage{amsfonts}                            % Use AMS fonts
\usepackage{amssymb}                             % and symbols
\usepackage[mathscr]{eucal}                      % Euler font symbols
\usepackage{dcolumn}                             % Decimal-dot aligned
\usepackage{theorem}                             % Allow use of theorems
\usepackage{epsfig}                              % References as links
\usepackage{amsmath}                              % References as links
\usepackage{longtable}
\begin{document}

\begin{flushleft}
\vspace*{-0.99cm}
DESY 03-051\\
LU-ITP 2003/006\\
Edinburgh-2003/04\\
LTH-573\\
MIT-CTP-3365
\vspace*{-0.25cm}
\end{flushleft}
%\vspace*{-0.5cm}

\title{Generalized Parton Distributions from Lattice QCD}

\author{M. G\"ockeler$^{1,2}$, R. Horsley$^{3}$, D. Pleiter$^{4}$,
        P.E.L. Rakow$^{5}$, A. Sch\"afer$^2$, G. Schierholz$^{4,6}$ and 
%        P.E.L. Rakow$^{5}$, G. Schierholz$^{4,6}$ and 
        W. Schroers$^{7}$
\vspace {0.15cm}}

\affiliation{$^1$ Institut f\"ur Theoretische Physik, Universit\"at Leipzig, 
D-04109 Leipzig, Germany\\
$^2$ Institut f\"ur Theoretische Physik, Universit\"at Regensburg, 
D-93040 Regensburg, Germany\\
$^3$ School of Physics, University of Edinburgh, Edinburgh EH9 3JZ, UK\\
$^4$ John von Neumann-Institut f\"ur Computing NIC, Deutsches 
Elektronen-Synchrotron DESY, D-15738 Zeuthen, Germany\\
$^5$ Theoretical Physics Division, Department of Mathematical 
Sciences, University of Liverpool, Liverpool L69 3BX, UK\\
$^6$ Deutsches Elektronen-Synchrotron DESY, D-22603 Hamburg, Germany\\
$^7$ Center for Theoretical Physics, Massachusetts Institute of 
Technology, Cambridge, MA 02139, USA}
\author{\vspace{-0.3cm} - QCDSF Collaboration -}

\begin{abstract}
  We perform a quenched lattice calculation of the first moment of twist-two
  generalized parton distribution functions of the proton, and assess the 
  total 
  quark (spin and orbital angular momentum) contribution to the spin of the 
  proton. 
\end{abstract}


%\keywords{Generalized parton distribution, proton, total quark angular
%momentum} 

\maketitle

Generalized parton distributions~\cite{GPD} (GPDs) provide a deeper
understanding of the internal structure of hadrons in terms of quarks and
gluons. While ordinary parton distributions measure the probability
$|\psi(x)|^2$ of finding a parton with fractional momentum $x$ in the hadron,
GPDs describe the coherence of two different hadron wave functions
$\psi^\dagger(x+\xi/2)\,\psi(x-\xi/2)$, one where the parton carries fractional
momentum $x+\xi/2$ and one where this fraction is $x-\xi/2$, from which
information about parton-parton correlation functions can be deduced. As a
consequence, GPDs
depend on the momentum transfer $\Delta^2$ between the initial and final 
hadron, which
provides further information on the transverse location of quarks and
gluons~\cite{Diehl}. Spatial images of hadrons can thus be obtained, where the 
resolution is determined by the virtuality $Q^2$ of the incoming photon.
Last, but not least, GPDs allow us to isolate the contribution of the quark 
orbital angular momentum to the spin of hadrons.
Lattice QCD is the only known method that is able to compute moments of GPDs
from first principles.

We will restrict ourselves to the GPDs $H_q$ and $E_q$ of the nucleon, 
where $q = u, d, \cdots$ denotes the flavor of the struck quark. We will not
consider the gluon sector here. The lowest, zeroth moments of $H_q$ and 
$E_q$ are given by the Dirac and Pauli form factors:  
\begin{eqnarray}
  \int_{-1}^1 \mbox{d}x\, H_q(x,\xi,\Delta^2) &=& F^q_1(\Delta^2) \,, \\
  \int_{-1}^1 \mbox{d}x\, E_q(x,\xi,\Delta^2) &=& F^q_2(\Delta^2) \,.
\end{eqnarray}
Both form factors have been computed on the lattice in a similar
calculation~\cite{QCDSF} to the present one and found to be well described
by a dipole ansatz
\begin{equation}
F_{1,2}^q(\Delta^2) = F_{1,2}^q(0)/(1-\Delta^2/M_{1,2}^2)^2
\label{df}
\end{equation}
for sufficiently small (and accessible) momenta,  
with dipole masses $M_{1,2}$ of the order of the $\rho$, $\omega$ mass, when
extrapolated to the physical pion mass. 

\begin{table*}[tbh]
\caption{Parameters of the dipole fit. In the bottom row we give the parameters
extrapolated to the physical pion mass.}
\begin{ruledtabular}
%\begin{tabular*}{20cm}{c|c|c|c}
\begin{tabular}{*8{c}}
$\kappa$ & $M\, \mbox{[GeV]}$ & $A_2^u(0)$ & $B_2^u(0)$ & $C_2^u(0)$ & 
$A_2^d(0)$ &
$B_2^d(0)$ & $C_2^d(0)$ \\ \hline
0.1324 & 1.69(05) & 0.419(07) & 0.344(028) & -0.084(26) & 0.188(04) & 
-0.281(20) & -0.071(15) \\
0.1333 & 1.58(06) & 0.415(10) & 0.334(044) & -0.101(35) & 0.176(05) &
-0.260(29) & -0.073(19) \\
0.1342 & 1.41(10) & 0.404(19) & 0.357(117) & -0.117(70) & 0.158(10) &
-0.265(80) & -0.067(35) \\ \hline
 & 1.11(20) & 0.400(22) & 0.334(113) & -0.134(81) & 0.147(11) & 
-0.232(77) & -0.071(42)\\
\end{tabular}
\end{ruledtabular}
\end{table*}

The first moments of $H_q$ and $E_q$ are of the form~\cite{GPD} 
\begin{eqnarray}
  \int_{-1}^1 \mbox{d}x\, x\, H_q(x,\xi,\Delta^2) &=& A_2^q(\Delta^2) + 
  \xi^2 C_2^q(\Delta^2)\,,\label{m1}\\
  \int_{-1}^1 \mbox{d}x\, x\, E_q(x,\xi,\Delta^2) &=& B_2^q(\Delta^2) - 
  \xi^2 C_2^q(\Delta^2)\,,\label{m2}
\end{eqnarray}
where $A_2^q(\Delta^2)$, $B_2^q(\Delta^2)$ and 
$C_2^q(\Delta^2)$ are generalized form factors, which are given by the nucleon 
matrix elements of the energy-momentum tensor (EMT):
\begin{equation}
\begin{split}
  \langle p'|{\cal O}^q_{\lbrace\mu\nu\rbrace}| p\rangle &\equiv
  \frac{\mbox{i}}{2} \langle p'|\bar{q}\gamma_{\lbrace\mu} 
  {\stackrel{\leftrightarrow}{D}}_{\nu\rbrace} q | p\rangle \\
   &= A_2^q(\Delta^2)\, \bar{u}(p')\gamma_{\lbrace\mu}\bar{p}_{\nu\rbrace} u(p)
  \\ 
   &- B_2^q(\Delta^2)\,\frac{\mbox{i}}{2m_N}\bar{u}(p')\Delta^\alpha
  \sigma_{\alpha\lbrace\mu}\bar{p}_{\nu\rbrace} u(p)  \\    
   &+ C_2^q(\Delta^2)\,\frac{1}{m_N}\bar{u}(p')u(p)\Delta_{\lbrace\mu}
  \Delta_{\nu\rbrace}\,.
\label{me}
\end{split}
\end{equation}
Here $m_N$ denotes the nucleon mass, $\bar{p}=\frac{1}{2}\left(p'+p\right)$, 
$\Delta=p'-p$, and curly brackets refer to symmetrization of indices and
subtraction of traces. The EMT has twist two and spin two. It is assumed to be
renormalized at the scale $\mu$, which makes $A_2^q(\Delta^2)$, 
$B_2^q(\Delta^2)$ and $C_2^q(\Delta^2)$
scale and scheme dependent. For the classification of states of definite
$J^{PC}$ contributing to (\ref{me}) in the $t$-channel see \cite{Ji&Lebed}. 
The so-called skewedness parameter $\xi$ is 
defined by $\xi = -n\cdot\Delta$, where $n$ is a light-like vector with 
$n\cdot\bar{p}=1$, and bounded by $|\xi| \leq
2\sqrt{\Delta^2/(\Delta^2-4m_N^2)}$. 
%The generalized form factors $A_2^q(\Delta^2)$, $B_2^q(\Delta^2)$ and 
%$C_2^q(\Delta^2)$ are real,
%non-singular functions for $\Delta^2 \leq 0$. 
In the forward limit, $\Delta^2 \rightarrow 0$, we have
\begin{equation}
A_2^q(0) = \langle x_q\rangle 
\equiv \int_0^1 \mbox{d}x\, x\, \big(q_\uparrow(x)+q_\downarrow(x)\big)\,,
\end{equation}
where $q_{\uparrow(\downarrow)} (x)$ are the usual quark distributions with
spin parallel (antiparallel) to the spin of the nucleon. Furthermore, one
derives~\cite{Ji}
\begin{equation}
\frac{1}{2} \big(A_2^q(0)+B_2^q(0)\big) = J_q \,,
\label{J}
\end{equation}
where $J_q$ is the angular momentum of the $q$ quark, and
%\begin{equation}
$J = \sum_q J_q$
%\end{equation}
is the total angular momentum of the nucleon carried by the quarks. The 
angular momentum decomposes,
in a gauge invariant way, into two pieces:
\begin{equation}
J_q = L_q + S_q \,,
\end{equation}
where $L_q$ is the orbital angular momentum and
\begin{equation}
S_q = \frac{1}{2} \Delta q \equiv \frac{1}{2} \int_0^1 \mbox{d}x\,
\big(q_\uparrow(x) - q_\downarrow(x)\big)
\end{equation}
is the spin of the quark. 
We know $\Delta q$ from separate calculations~\cite{QCDSF3,Deltaq}, 
so that $L_q$ can be computed from (\ref{J}).

In this Letter we perform a quenched lattice calculation of the 
generalized form factors $A_2^q(\Delta^2)$, $B_2^q(\Delta^2)$ and 
$C_2^q(\Delta^2)$. The quenched approximation neglects fluctuations of
virtual quark-antiquark pairs from the Dirac sea. 
%Their effect on the form factors is expected to be 
%small, as can be drawn from our calculation of $F_{1,2}$~\cite{QCDSF}. 
The non-forward matrix elements (\ref{me}) are 
computed from ratios of three- and two-point functions following~\cite{QCDSF}.
Further details are given in~\cite{MIT}. To keep cut-off effects 
small, we use non-perturbatively $\mathcal{O}(a)$ improved Wilson fermions.
The calculation is done on $16^3\,32$ lattices at $\beta=6.0$ and for three 
different hopping parameters, $\kappa=0.1324$, $0.1333$ and 
$0.1342$, which allows us to extrapolate our results to the chiral limit. 
Using $r_0=0.5\,\mbox{fm}$ to set the scale, which results in the inverse 
lattice spacing $1/a=2.12\,\mbox{GeV}$, the corresponding 
pion masses are $1070$, $870$ and 
$640\,\mbox{MeV}$. If we use the nucleon mass extrapolated to the chiral 
limit to set the scale, the pion masses are $930$, $760$ and 
$550\,\mbox{MeV}$, and $1/a=1.84\,\mbox{GeV}$. 
The corresponding nucleon masses and the choice of nucleon momenta $p, p'$ can
be inferred from~\cite{QCDSF}. 
For the EMT we consider two sets of (euclidean) operators:
\begin{equation}
\frac{1}{\sqrt{2}}\big(\mathcal{O}_{\mu\nu}+\mathcal{O}_{\nu\mu}\big)\,, 
\; 1 \leq \mu < \nu \leq 4
\label{o1}
\end{equation}
and
\begin{equation}
\begin{tabular}{c}
$\displaystyle 
\frac{1}{2}\big(\mathcal{O}_{11}+\mathcal{O}_{22}-\mathcal{O}_{33}-
\mathcal{O}_{44}\big)$\,, \\[1em]
$\displaystyle 
\frac{1}{\sqrt{2}}\big(\mathcal{O}_{33}-\mathcal{O}_{44}\big)\,, \,
\frac{1}{\sqrt{2}}\big(\mathcal{O}_{11}-\mathcal{O}_{22}\big)$\,.
\end{tabular}
\label{o2}
\end{equation}
Each set transforms irreducibly under the hypercubic group.
The operators (\ref{o1}) and (\ref{o2}) are renormalized multiplicatively, 
%\begin{equation}
$\mathcal{O}(\mu) = Z(a\mu)\, \mathcal{O}(a)$,  
%\end{equation}
with renormalization constants~\cite{QCDSF3} $Z_{v_{2a}}$ and
$Z_{v_{2b}}$, respectively. 
The renormalization constants %$Z_{v_{2a}}$ and $Z_{v_{2b}}$
are computed non-perturbatively~\cite{QCDSF4} following~\cite{Z}. 
We obtain $Z^{\overline{MS}}_{v_{2a}}(2\,\mbox{GeV}) = 
1.10$ and $Z^{\overline{MS}}_{v_{2b}}(2\,\mbox{GeV}) = 1.09$. The following 
results refer to the $\overline{MS}$ scheme at the renormalization scale
$\mu = 2\,\mbox{GeV}$.

\begin{figure}[b,h,t]
%\caption{The generalized form factors $A_u$, $B_u$ and $C_u$ on the $16^3\,32$
%  lattice at $\beta=6.0,
%  \kappa=0.133$ together with a
%dipole fit.}
\begin{centering}
\epsfig{figure=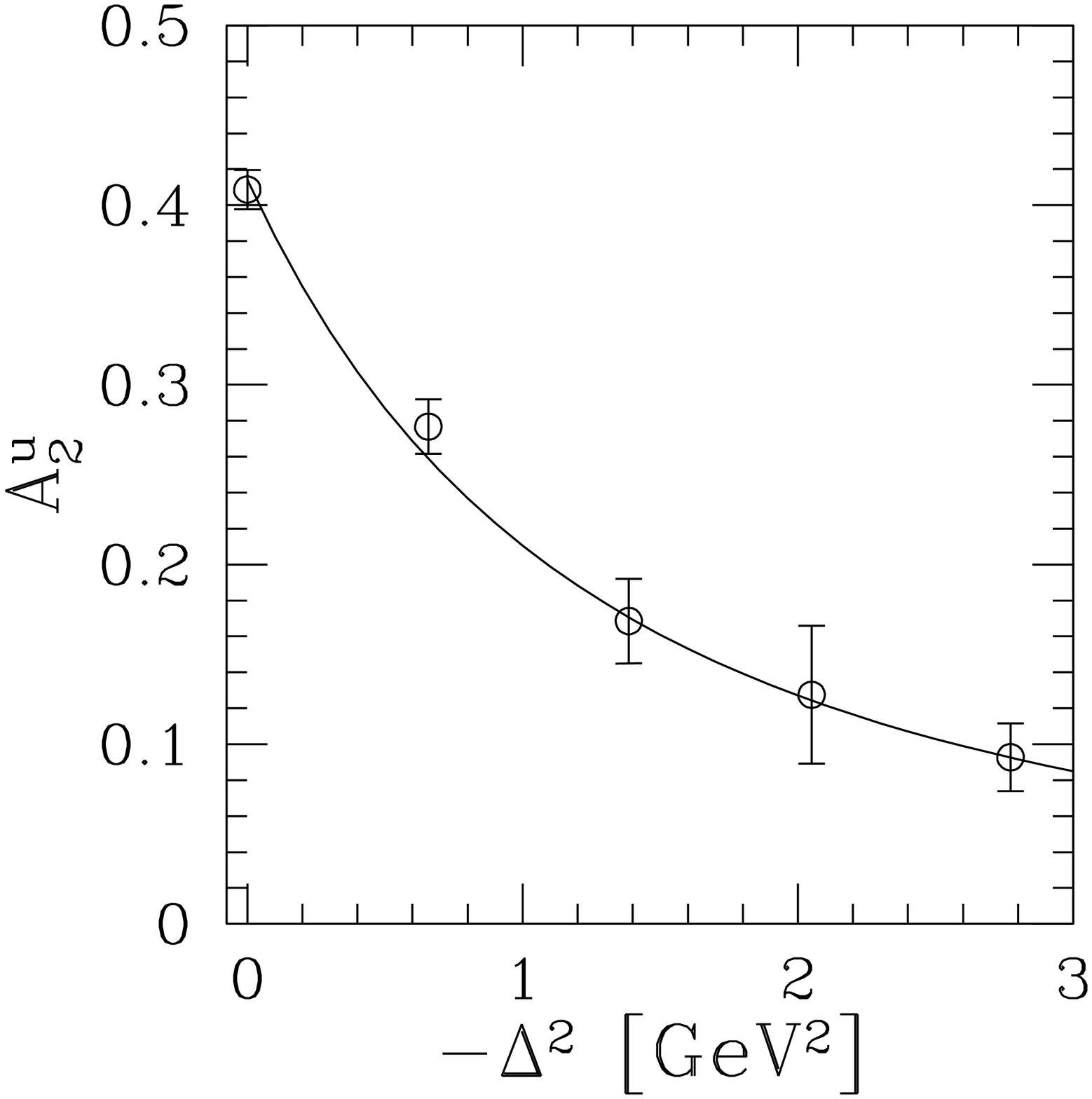,height=7cm,width=7cm}\\
\epsfig{figure=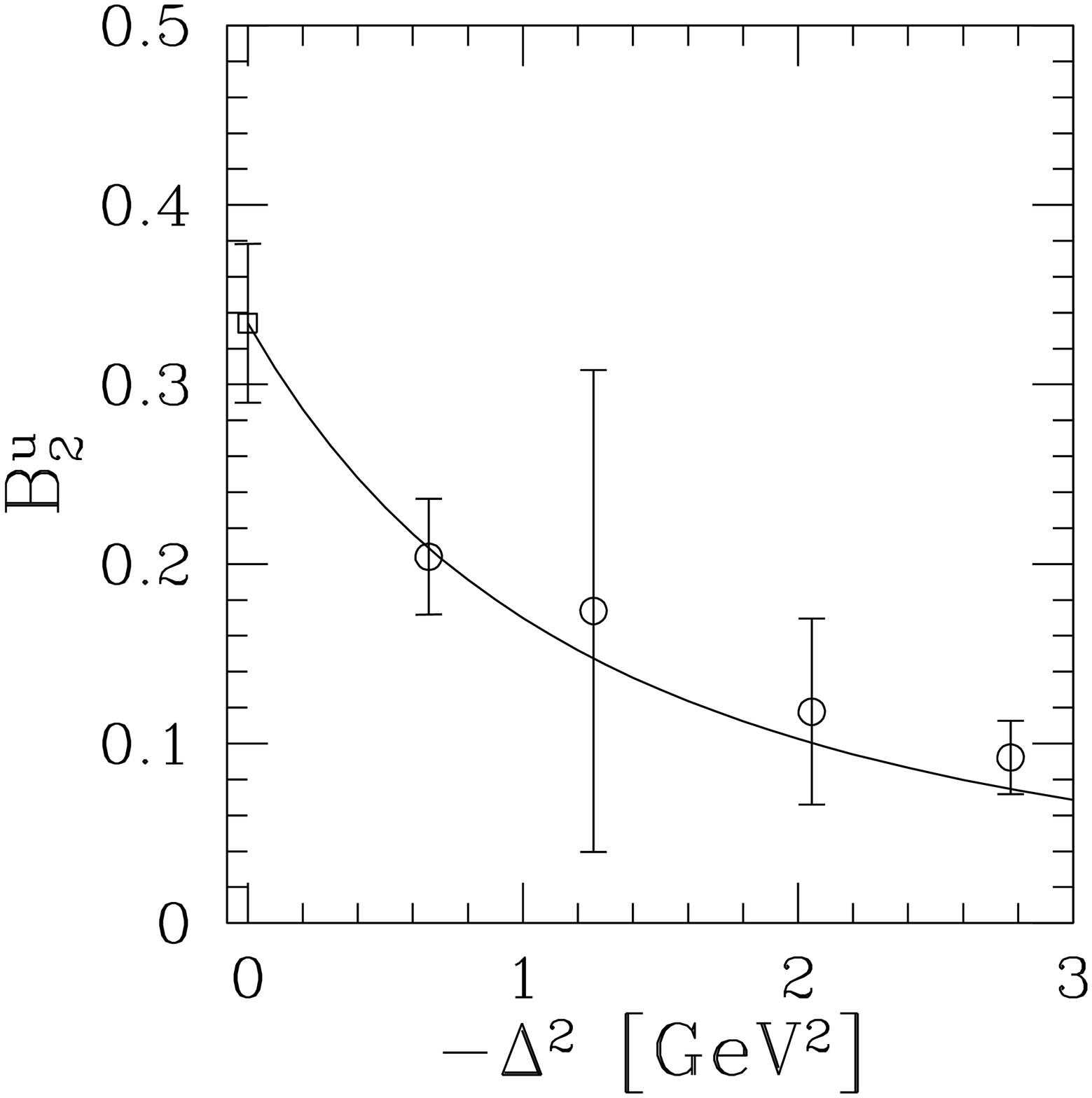,height=7cm,width=7cm}\\
\epsfig{figure=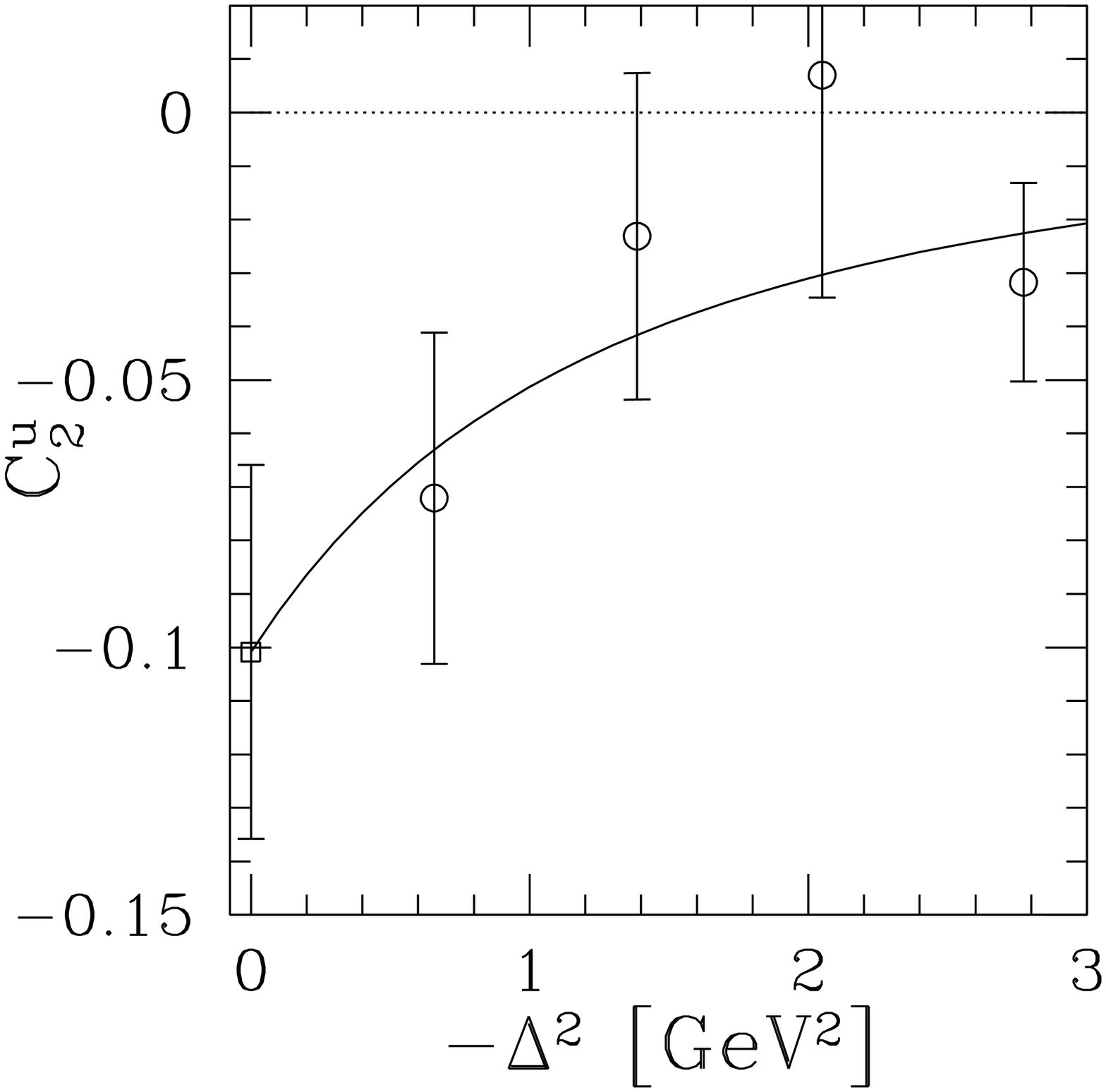,height=7cm,width=7cm}
%\caption{ } 
\end{centering}
\caption{The generalized form factors $A_2^u$, $B_2^u$ and $C_2^u$ at 
$\kappa=0.1333$, together with the dipole fit and the extrapolated values 
at $\Delta^2=0$ ($\Box$).}
%\vspace*{-0.15cm}
\end{figure}

\begin{figure}[b,h,t]
\begin{centering}
\epsfig{figure=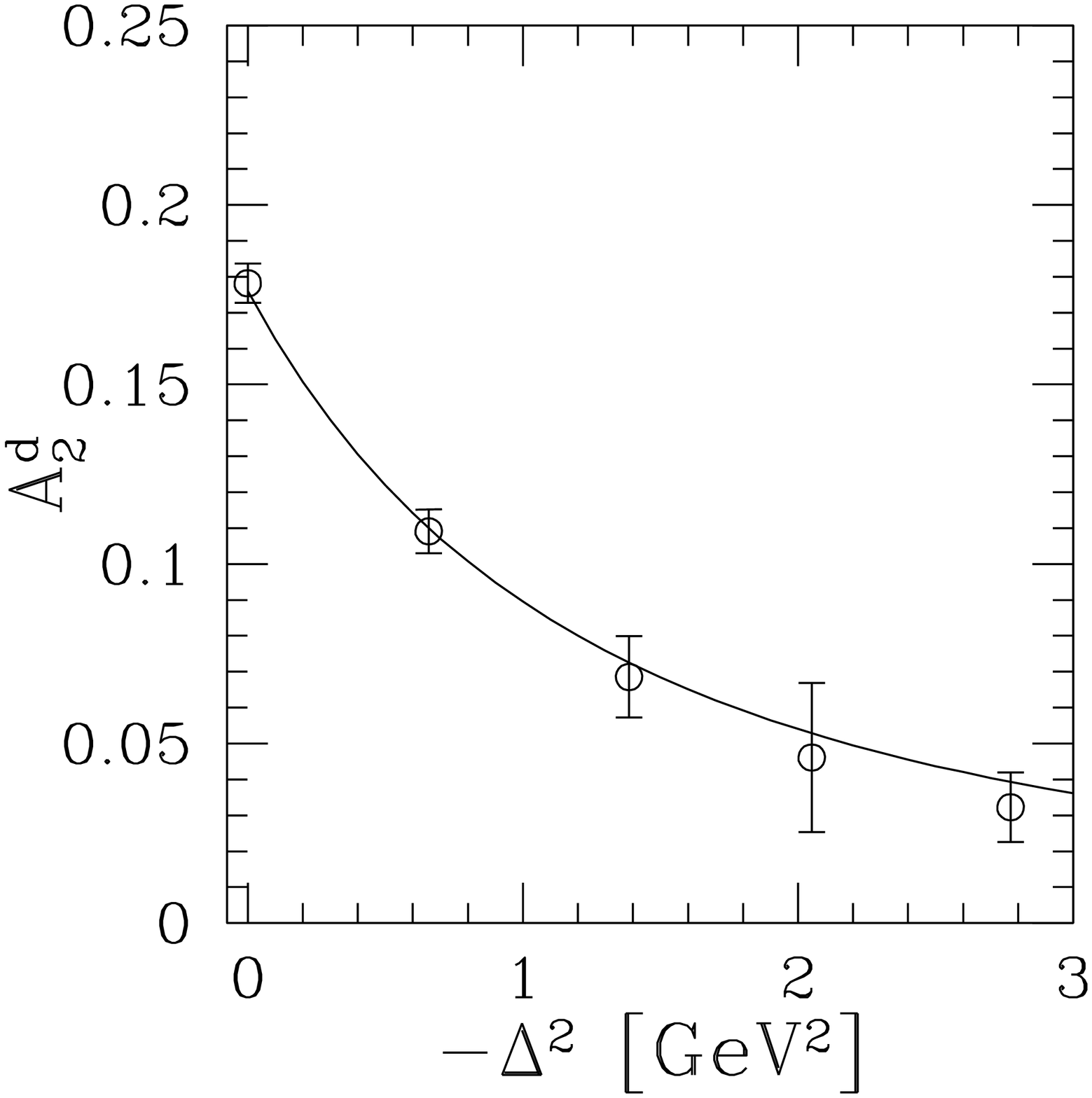,height=7cm,width=7cm}\\
\epsfig{figure=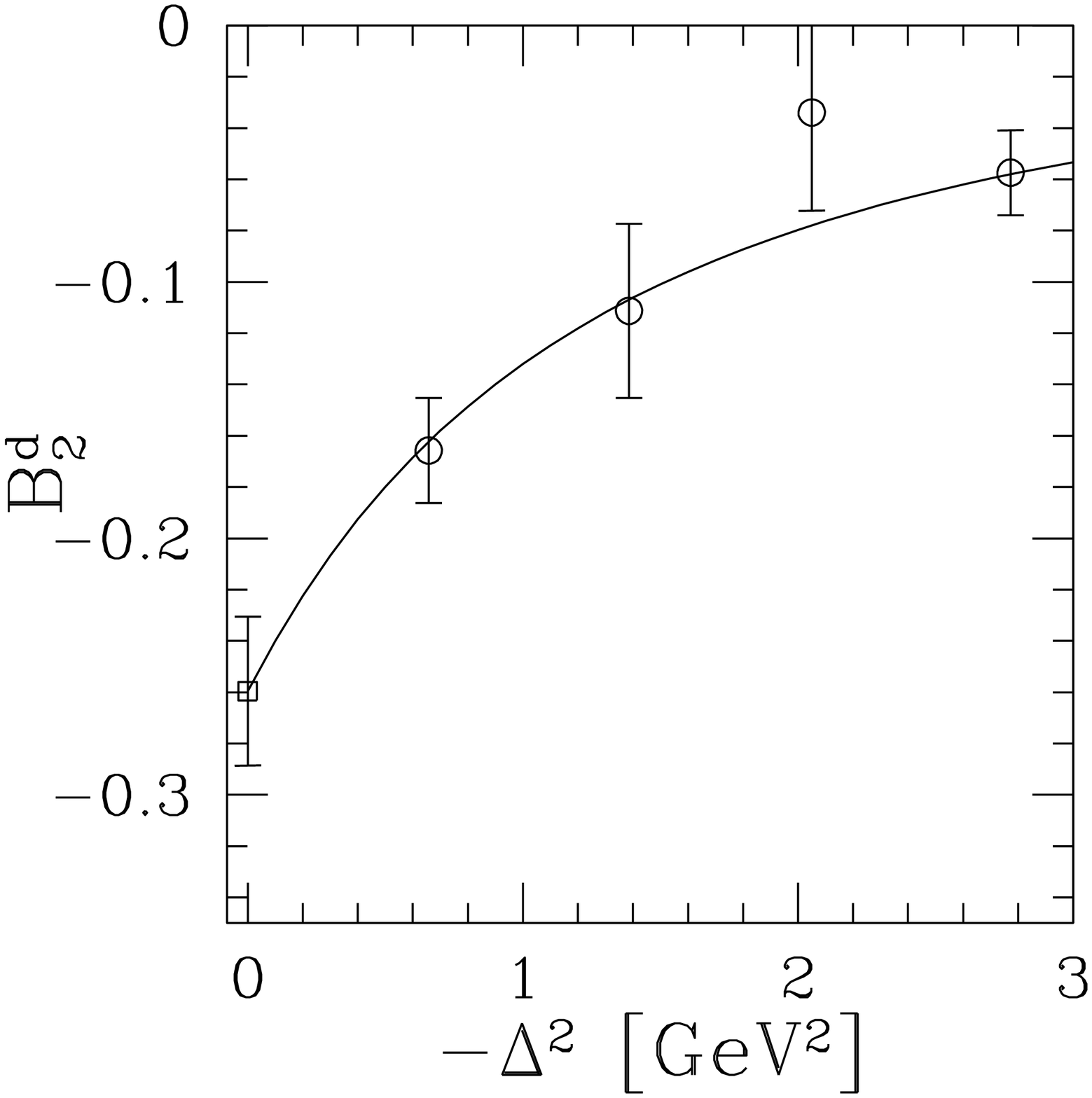,height=7cm,width=7cm}\\
\epsfig{figure=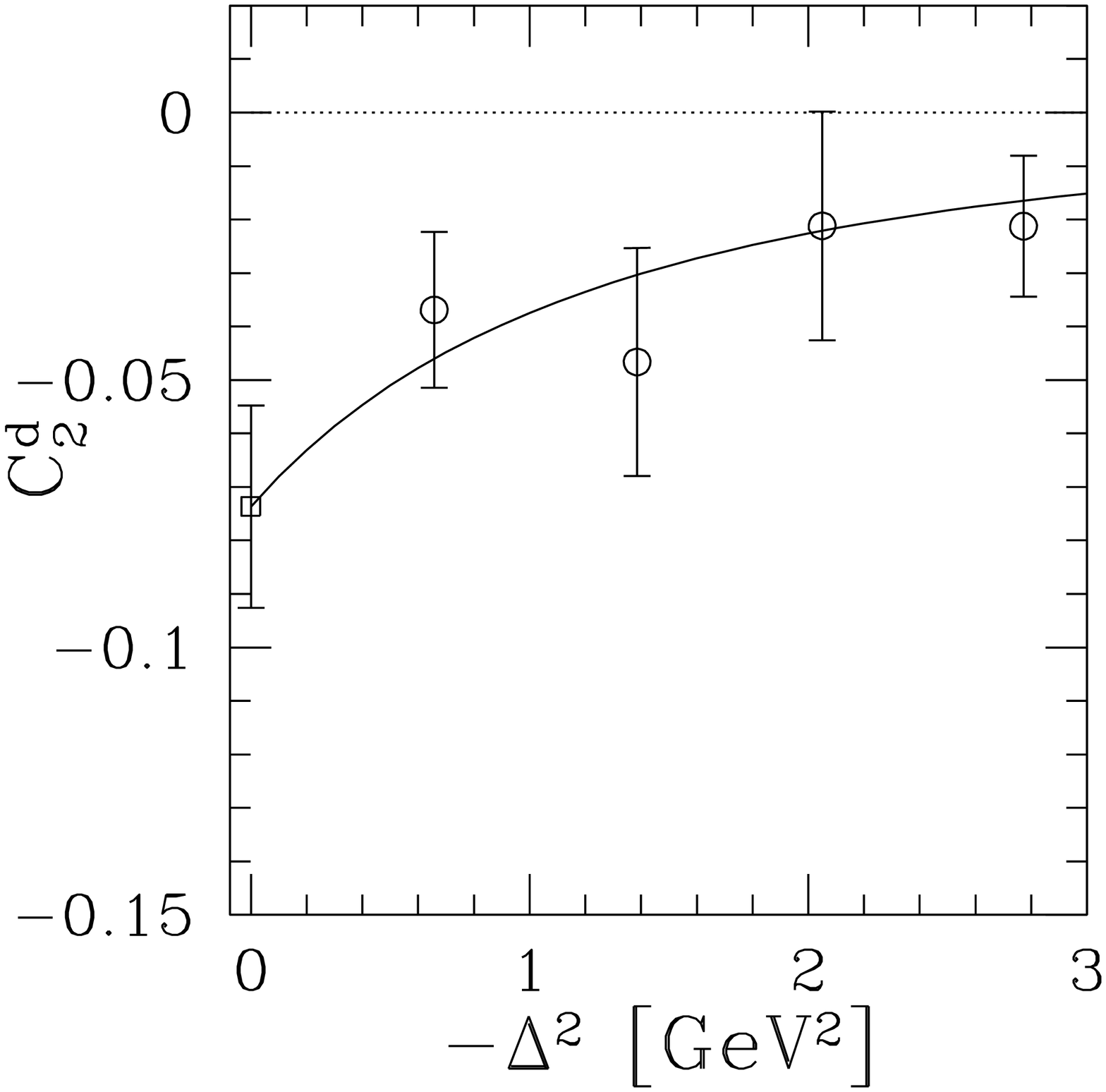,height=7cm,width=7cm}
%\caption{ } 
\end{centering}
\caption{The generalized form factors $A_2^d$, $B_2^d$ and $C_2^d$ at 
$\kappa=0.1333$, together with the dipole fit and the extrapolated values 
at $\Delta^2=0$ ($\Box$).}
%\vspace*{-0.35cm}
\end{figure}

In Figs.~1 and 2 we show the generalized form factors 
$A_2^u(\Delta^2)$, $B_2^u(\Delta^2)$, $C_2^u(\Delta^2)$ and
$A_2^d(\Delta^2)$, $B_2^d(\Delta^2)$, $C_2^d(\Delta^2)$ of the proton for
$\kappa = 0.1333$. Data points with larger errors are not shown here but are 
included in the fit. The corresponding form factors of the neutron are 
obtained by interchanging $u$ and $d$. Similarly good results are found for 
$\kappa=0.1324$ and $0.1342$. The generalized form factors can be well 
described by the dipole ansatz
\begin{equation}
A_2^q(\Delta^2) = A_2^q(0)/(1-\Delta^2/M^2)^2 \,,
\label{dgpd}
\end{equation}
and similarly for $B_2^q$ and $C_2^q$. Fits of $A_2^u(\Delta^2)$ and 
$A_2^d(\Delta^2)$
give the same dipole mass $M$ within errors. The dipole masses obtained from 
separate fits of $B_2^u(\Delta^2)$, $B_2^d(\Delta^2)$, $C_2^u(\Delta^2)$ and 
$C_2^d(\Delta^2)$ are found to be consistent with that value. We therefore have
decided to fit our data by a common dipole mass $M$. Our data do not favor
a monopole behavior. The results of the fits are shown in Table I. For a 
reliable
extrapolation to $\Delta^2 = 0$ we find it important to cover a wide enough
range of $\Delta^2$ values. This may be the reason why our dipole masses turn 
out to be systematically larger than those found in a previous 
calculation~\cite{Liu}.  
%Values for $B_q(0)$ and $C_q(0)$ are obtained by extrapolation.

%One expects the dipole mass to be of the order of the lightest tensor meson 
%mass. 
In Fig.~3 we show the dipole mass $M$ as a function of the pion mass. The 
mass values 
appear to lie on a straight line, as was observed already in the case of 
the nucleon form factors~\cite{QCDSF}. A linear extrapolation in $m_\pi$ to 
the physical pion mass gives $M = 1.1(2)\,\mbox{GeV}$. This value is close 
to the physical masses of the $f_2$, $a_2$ mesons, which supports the 
hypothesis of tensor meson dominance.
A quadratic extrapolation in $m_\pi$ leads to $M = 1.3(1)\,\mbox{GeV}$. The 
form factor
data $A_2^q(0)$, $B_2^q(0)$ and $C_2^q(0)$ show little variation with the 
quark mass and are extrapolated quadratically in $m_\pi$ to the physical pion 
mass. 
The results are shown in the bottom row of Table I. It should be stressed 
that all quantities refer (at best) to valence quark distributions, because 
sea quark effects have been neglected. 
In unquenched simulations there are also quark-line disconnected contributions.
For an estimate see~\cite{Liu}.

%We expect the dipole mass to be of the order of the $f_2\,(1270)$, 
%$a_2\,(1320)$ mass, because the EMT has spin two. In Fig.~2 we show $M$ as a 
%function of the pion mass. The fitted values lie on a straight line. A 
%linear extrapolation to the physical pion mass gives $M = 1.1(2)\,\mbox{GeV}$,
%which is in reasonable agreement with the lightest tensor meson mass. A 
%quadratic extrapolation in $m_\pi$ leads to a similar result. The numbers 
%$A_q(0)$, $B_q(0)$ and $C_q(0)$ appear to be independent of the quark mass 
%within the errors. It should be stressed that all quantities refer (at best)
%to valence quark distributions because effects of sea quarks have been 
%neglected.

\begin{figure}[t!]
\begin{centering}
\epsfig{figure=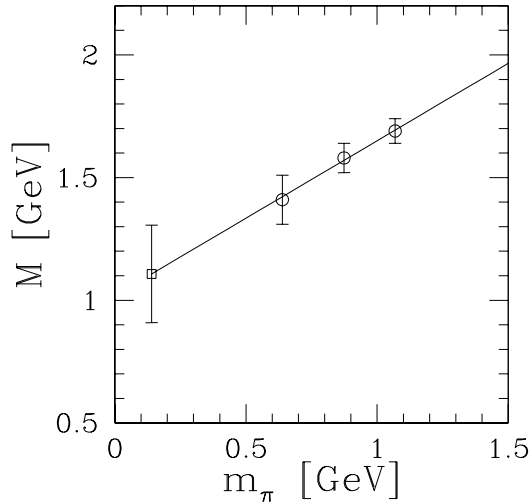,height=7cm,width=7cm}
\caption{The dipole mass $M$ as a function of $m_\pi$, together with a linear
extrapolation to the physical pion mass ($\Box$).} 
\end{centering}
%\vspace*{-0.1cm}
\end{figure}

If the dipole behavior (\ref{df}), (\ref{dgpd}) continues to hold for the 
higher moments as well, and if we assume that the dipole masses 
continue to grow in a Regge-like fashion, we would obtain
\begin{equation} 
\int_{-1}^1 \mbox{d}x\, x^n \,H_q(x,0,\Delta^2) = 
\langle x_q^n\rangle /(1-\Delta^2/M_{n+1}^2)^2 \,,
\label{H}
\end{equation}
%($\langle x_q^0\rangle = F_1^q(0)$) 
with $M_l^2 = \mbox{const.} + l/\alpha'$, $\alpha'$ being the slope of the 
Regge trajectory. 
This would mean that with increasing momentum transfer $|\Delta^2|$ the 
lower moments of $H_q(x,0,\Delta^2)$ are suppressed more than the higher ones,
so that the observed peak in $H_q(x,0,0) = q_\uparrow(x)+q_\downarrow(x)$ 
around $x \approx 0.2$ is shifted towards the higher values of $x$. As a 
result, the $\Delta^2$ dependence cannot be factorized in a simple way, as is 
sometimes assumed. Knowing $\langle x_q^n\rangle$, we can reconstruct
$H_q(x,0,\Delta^2)$ from (\ref{H}) by inverse Mellin transform. 
The $\xi$ dependence of both $H_q$ and $E_q$ appears to be rather weak, based 
on our knowledge of the first two moments, and in 
the isovector channel (corresponding to proton--neutron or $u$--$d$ matrix 
elements) it largely cancels out.

In Fig.~4 we show the total angular momentum $J=J_u+J_d$. The dependence on 
the pion mass is
rather flat, as expected~\cite{Chen}. The errors are due to the relatively 
large statistical errors of 
$B_2^u$ and $B_2^d$ and the fact that $B_2^u$ and $B_2^d$ cancel to a large 
extent. 
In Table II we give our results for $J$, and separately for $J_q$ and $S_q$,
extrapolated quadratically (linearly in $m_\pi^2$) to the physical pion mass.
The numbers for $S_q$ 
refer to our latest results~\cite{QCDSF4}, computed from the non-perturbatively
improved axial vector current with non-perturbative renormalization 
factors. It turns out that the total angular momentum $J$ carried by the 
quarks amounts to 
$\approx 70\%$ of the spin of the (quenched) proton, leaving a contribution
of $\approx 30\%$ for the gluons. 
%(Note again that sea quarks are absent in our calculation.) 
The major contribution is given by the $u$ quark, while the contribution of 
the $d$ quark is found to be negligible, which hints at strong pairing 
effects. Our result for $J$ is somewhat smaller than that of 
\cite{Liu,Gadiyak}.
We are able to compute $L_q$ now. The total 
orbital angular momentum of the quarks turns out to be consistent 
with zero:
\begin{equation}
L \equiv L_u + L_d = 0.03(7).  
\end{equation}
This indicates that (at virtuality $\mu = 2\,\mbox{GeV}$) the parton's 
transverse momentum in the (quenched) proton is small. A similar conclusion 
can be drawn from our earlier finding~\cite{d2} 
of a small twist-three contribution $d_2$ to the second moment of the polarized
structure function $g_2$.  

\begin{figure}[t!]
\begin{centering}
\epsfig{figure=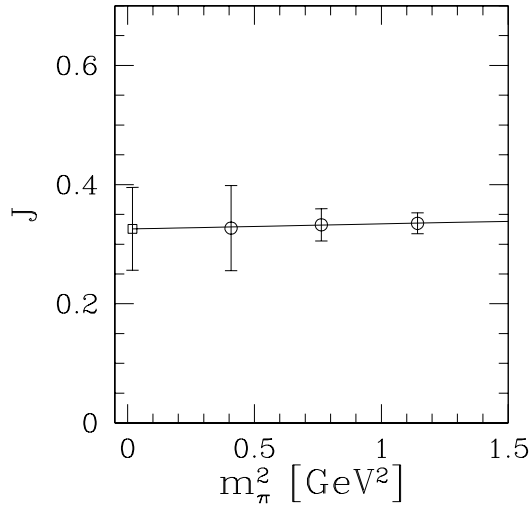,height=7cm,width=7cm}
\caption{The total angular momentum $J$, together with a
quadratic extrapolation to the physical pion mass ($\Box$).} 
\end{centering}
%\vspace*{-0.15cm}
%\vspace*{-0.35cm}
%\vspace*{-0.5cm}
\end{figure}

\begin{table}
\caption{The total angular momentum and its individual contributions, 
extrapolated to the physical pion mass.}
\begin{ruledtabular}
\begin{tabular}{*{5}{c}}
$J$ & $J_u$ & $J_d$ & $S_u$ & $S_d$ \\
\hline
0.33(7) & 0.37(6) & -0.04(4) & 0.42(1) & -0.12(1) \\
\end{tabular}
\end{ruledtabular}
\vspace*{-0.15cm}
\end{table}

The generalized form factors $C_2^q(\Delta^2)$ contribute to the beam charge
asymmetry of deeply virtual Compton scattering. We obtain a rather
small value: $C_2^u(0) + C_2^d(0) = -0.2(1)$. This result is to be compared 
with the value $-0.8$ obtained in the chiral quark-soliton model at 
$\mu \approx 0.6 \,\mbox{GeV}$~\cite{Kivel}. For a discussion
see also~\cite{pheno}. 

As far as one can compare, quenched and unquenched results agree surprisingly
well, and we do not expect to find significant differences here either. 
For a recent study of quenching artifacts, as well as cut-off effects,
see~\cite{QCDSF5}. 

\begin{acknowledgments}
This work is supported by DFG and under the Feodor Lynen program. Discussions 
with P.~H{\"a}gler, J.~Negele, D.~Renner and C.~Weiss are acknowledged. The
numerical calculations have been performed at NIC (J\"ulich).
\end{acknowledgments}

\vspace*{0.9cm}
\end{document}